\documentclass[a4paper]{article}
\usepackage{graphicx}
\usepackage{amsmath}
\usepackage{amsfonts}
\usepackage{amssymb}
\newcommand{\be}{\begin{equation}}
\newcommand{\ee}{\end{equation}}

\newcommand{\al}{\alpha}

\newcommand{\om}{\omega}

\newcommand{\G}{\Gamma}

\newcommand{\dz}{\wedge}
\newcommand{\C}{{\bf C }}

\newcommand{\ba}{\begin{array}}
\newcommand{\ea}{\end{array}}
\newcommand{\beq}{\begin{eqnarray}}
\newcommand{\eeq}{\end{eqnarray}}
\newtheorem{lm}{Lemma}
\newtheorem{thee}{Theorem}
\newtheorem{proo}{Proposition}
\newtheorem{co}{Corollary}
\newtheorem{rem}{Remark}
\newtheorem{deff}{Definition}
\newcommand{\bd}{\begin{deff}}
\newcommand{\ed}{\end{deff}}

\newcommand{\bl}{\begin{lm}}
\newcommand{\el}{\end{lm}}
\newcommand{\bp}{\begin{proo}}
\newcommand{\ep}{\end{proo}}
\newcommand{\bt}{\begin{thee}}
\newcommand{\et}{\end{thee}}
\newcommand{\bc}{\begin{co}}
\newcommand{\ec}{\end{co}}
\newcommand{\brm}{\begin{rem}}
\newcommand{\erm}{\end{rem}}

\newcommand{\der}{{\rm d}}
\begin{document}
\thispagestyle{empty}
\title{{Conformal Einstein equations and Cartan conformal connection} 
\vskip 1.truecm
\author{Carlos Kozameh\\
FaMAF Universidad Nacional de Cordoba\\
Ciudad Universitaria\\
Cordoba 5000\\
Argentina\\
\\
\\
Ezra T Newman\\
Department of Physics and Astronomy\\
University of Pittsburgh\\
Pittsburgh PA\\
USA\\
newman@pitt.edu\\
\\
\\
Pawel Nurowski\\
Instytut Fizyki Teoretycznej\\
Uniwersytet Warszawski\\
ul. Hoza 69, Warszawa\\
Poland\\
nurowski@fuw.edu.pl}
}

\author{\mbox{}}
\maketitle
\begin{abstract}
Necessary and sufficient conditions for a space-time to be conformal
to an Einstein space-time are interpreted in terms of curvature
restrictions for the corresponding Cartan conformal connection.
\end{abstract}

\newpage
\noindent

%\vglue 1.5truecm
\rm
\noindent
In Ref.\cite{CTP} we gave necessary
and sufficient conditions for a 4-dimensional metric to be conformal
to an Einstein metric. One of these conditions, the
vanishing of the Bach tensor of the metric, has been discussed by many
authors \cite{CTP,LK,Mas,Merk}. In particular, it was interpreted
as being equivalent to the vanishing of the Yang-Mills current of the
corresponding Cartan conformal connection. The other condition, which
is given in terms of rather complicated equation on the Weyl tensor of
the metric has not been 
analyzed from the point of view of the corresponding Cartan
conformal connection. The purpose of this letter is to fill this
gap.\\
  
\noindent
Let $M$ be a 4-dimensional manifold equipped with the conformal class
of metrics $[g]$. Here
we will be assuming that $g$ has Lorentzian signature, but our results
are also valid in the other two signatures.\\

\noindent
Given a conformal class $[g]$ on $M$ we choose a 
representative $g$ for the metric. Let $\theta^\mu$, $\mu=1,2,3,4$, 
be a null (or orthonormal) coframe for $g$ on $M$. This, in
particular, means that $g=\eta_{\mu\nu}\theta^\mu\theta^\nu$, with all the 
coefficients $\eta_{\mu\nu}$ being constants. We define $\eta^{\mu\nu}$ by 
$\eta^{\mu\nu}\eta_{\nu\rho}=\delta^\mu_{~\rho}$ and we will use 
$\eta^{\mu\nu}$ and 
$\eta_{\mu\nu}$ to rise and lower the Greek indices, respectively. The
metric 
\be
g'={\rm e}^{-2\phi}g\label{gprim}
\ee 
conformally related to $g$ will be 
represented by a coframe 
\be
\theta'^\mu={\rm e}^{-\phi}\theta^\mu
\ee
so that  
\be
g'=\eta_{\mu\nu}\theta'^\mu\theta'^\nu
\ee 
with the same $\eta_{\mu\nu}$ as in the expression (\ref{gprim}). 
% Let
%$e_\mu$ be the dual basis to $\theta^\mu$,
%$e_\nu\lrcorner\theta^\mu=\delta^\mu_{~\nu}$.
Given $g$ and $\theta^\mu$ we consider the 1-forms $\Gamma^\mu_{~\nu}$ 
uniquely determined on $M$ by the equations 
\beq
&\der\theta^\mu+\Gamma^\mu_{~\nu}\dz\theta^\nu=0,\nonumber\\
&\Gamma_{\mu\nu}=\Gamma_{[\mu\nu]},\quad\quad{\rm where}\quad\quad
\Gamma_{\mu\nu}=\eta_{\mu\rho}\Gamma^\rho_{~\nu}.
\eeq 
Using $\Gamma^\mu_{~\nu}$ we calculate the Riemann tensor 2-forms
\be
\Omega^\mu_{~\nu}=\frac{1}{2}\Omega^\mu_{~\nu\rho\sigma}\theta^\rho\dz\theta^\sigma=
\der\Gamma^\mu_{~\nu}+\Gamma^\mu_{~\rho}\dz\Gamma^{\rho}_{~\nu}, 
\ee
and the Ricci part of the Riemann tensor
\be
K_{\nu\sigma}=\Omega^\mu_{~\nu\mu\sigma},\quad\quad\quad 
K=\eta^{\nu\sigma}K_{\nu\sigma}\quad\quad\quad{\rm and}\quad\quad\quad
S_{\nu\sigma}=K_{\nu\sigma}-\frac{1}{4}K \eta_{\nu\sigma}.\label{ricci}
\ee
We recall that the metric $g$ is called {\it Einstein} iff
\be
S_{\mu\nu}=0
\ee
and that it is called {\it conformal to Einstein} iff there exists
$\phi$ such that the metric $g'={\rm e}^{-2\phi}g$ is Einstein.\\

\noindent 
In the following we will also need 1-forms
%\footnote{The antisymmetric
%part of $K_{\nu\sigma}$ is expressible in terms of
%$\Theta^\mu=\frac{1}{2}Q^\mu_{~\nu\rho}\theta^\nu\dz\theta^\rho$
%and its derivatives according to
%$K_{[\nu\sigma]}=-\frac{1}{2}\nabla_\mu
%Q^\mu_{~\nu\sigma}-\nabla_{[\nu}Q_{\sigma]}-\frac{1}{2}Q_\mu Q^\mu_{~\nu\sigma}$,
%where $Q_\mu=Q^\rho_{~\mu\rho}$ and $\nabla_\mu$ is the covariant
%derivative associated with $\Gamma^\mu_{~\nu}$.}
\be
\tau_\nu=\begin{pmatrix}
-\frac{1}{2}S_{\nu\rho}-\frac{1}{24}K \eta_{\nu\rho}
%-\frac{1}{n}K_{[\nu\rho]}
\end{pmatrix}\theta^\rho
\ee
and 2-forms
\be
C^\mu_{~\nu}=\der\Gamma^\mu_{~\nu}+\theta^\mu\dz\tau_\nu+\Gamma^\mu_{~\rho}\dz\Gamma^\rho_{~\nu}+\tau^\mu\dz\theta_\nu,~~~~~~\tau^\mu=\eta^{\mu\nu}\tau_\nu.
\ee
It follows that the 2-forms 
\be
C^\mu_{~\nu}=\frac{1}{2}C^\mu_{~\nu\rho\sigma}\theta^\rho\dz\theta^\sigma
\ee 
are the Weyl 2-forms 
associated with the Weyl tensor $C^\mu_{~\nu\rho\sigma}$ of the metric
$g$.  It is known that the Weyl tensor obeys the following identity
\be
C_{\al\mu\nu\rho}C^{\beta\mu\nu\rho}=\frac{1}{4} C^2\delta^\beta_{~\alpha},\label{wid}
\ee
where
\be
C^2=C_{\mu\nu\rho\sigma}C^{\mu\nu\rho\sigma}.
\ee
Using the Bianchi identities one shows that the 
tensor $\tau_{\mu\nu\rho}$ defined on $M$ via
\be
D\tau_\mu=\frac{1}{2}\tau_{\mu\nu\rho}\theta^\nu\dz\theta^\rho=\der\tau_\mu+\tau_\nu\dz\Gamma^\nu_{~\mu}
\ee 
is 
\be
\tau_{\nu\rho\sigma}=\nabla_\mu C^\mu_{~\nu\rho\sigma},\label{for}
\ee 
where $\nabla_\mu$ is the covariant derivative operator associated
with $D$.\\
%$D(T^{\al_1\al_2...\al_r}_{~~~~~~~~~~\bet_1\bet_2...\bet_s}\theta^{\bet_1}\dz\theta^{\bet_2}\dz...\dz\theta^{\bet_s})=\nabla_\mu
%(T^{\al_1\al_2...\al_r}_{~~~~~~~~~~\bet_1\bet_2...\bet_s})\theta^\mu\dz\theta^{\bet_1}\dz\theta^{\bet_2}\dz...\dz\theta^{\bet_s}$.\\

\noindent
In the following we will also need the Bach tensor 
\be
B_{\mu\nu}=\nabla^\rho\nabla^\sigma
C_{\mu\rho\nu\sigma}+\frac{1}{2}C_{\mu\rho\nu\sigma}K^{\rho\sigma}\label{bach}
\ee
and a tensor
\be
N_{\nu\rho\sigma}=(\nabla_\alpha
C^\alpha_{~\beta\gamma\delta})C^{\mu\beta\gamma\delta}C_{\mu\nu\rho\sigma}-\frac{1}{4} 
C^2 \nabla_\mu C^\mu_{~\nu\rho\sigma}\label{ted}.
\ee
It is known that the vanishing of the Bach tensor is a conformally 
invariant property. If $C^2\neq0$ the vanishing of $N_{\mu\nu\rho}$
is also conformally invariant. The relevance of both these tensors in
the context of this letter is given by the following theorem \cite{CTP}.
\bt
Assume that the metric $g$ satisfies the generiticity
condition 
$$C^2\neq 0$$ 
on $M$. Then the
metric is locally conformally equivalent to the Einstein metric if and
only if
\be
{\rm (i)}~~B_{\mu\nu}= 0~~~~{\rm and}~~~~{\rm (ii)}~~N_{\nu\rho\sigma}= 0
\ee
on $M$.
\et

\noindent
{\bf Remarks}
\begin{itemize}
\item Conditions (i) and (ii) are independent. In particular,
metrics with vanishing Bach tensor and not conformal to Einstein
metrics are known \cite{PN}.
\item If $C^2= 0$ the condition (ii) must be replaced by another
  condition for the above theorem to be true. This another condition depends
  on the algebraic type of the Weyl tensor and is given in
  \cite{CTP}. 
\item Baston and Mason \cite{Mas} gave another version of the above
  theorem in which condition (ii) was replaced by the vanishing of
  a different tensor than $N_{\nu\rho\sigma}$. Unlike
  $N_{\nu\rho\sigma}$, which is {\it cubic} in the Weyl tensor, the
  Baston-Mason tensor $E_{\nu\rho\sigma}$, is
  only {\it quadratic} in $C^\mu_{~\nu\rho\sigma}$.
\item Merkulov \cite{Merk} interpreted condition (ii) as the vanishing
  of the Yang-Mills current of the {\it Cartan
  normal conformal connection} $\om$ associated with the metric
  $g$. Following him, Baston and Mason \cite{Mas} interpreted the
  condition $E_{\nu\rho\sigma}=0$ in terms of curvature condition for $\om$.
\end{itemize}

\noindent
Although in the context of Theorem 1 
conditions (ii) and $E_{\nu\rho\sigma}=0$ are equivalent, the tensors
$N_{\mu\nu\rho}$ and $E_{\mu\nu\rho}$ are quite different. In addition
to cubic versus quadratic dependence on the Weyl tensor, one can
mention the fact that it is quite easy to express tensor
$E_{\nu\rho\sigma}$ in the spinorial language and quite complicated in
the tensorial language. Totally oposite situation occurs for the
tensor $N_{\nu\rho\sigma}$. One of the motivation for the present
letter is the existence of the normal conformal connection interpretation for
the condition $E_{\nu\rho\sigma}=0$. As far as we know such
interpretation of $N_{\nu\rho\sigma}=0$ has not been discussed. To
fill this gap we first give the formal definition of the Cartan normal
conformal connection. In order to do this we first, introduce the 
$6\times 6$ matrix
\be
Q_{AB}=\begin{pmatrix}
0&0&-1\\
0&\eta_{\mu\nu}&0\\
-1&0&0
\end{pmatrix}
\ee
and then define the ${\bf so}(2,4)$-valued 1-form 
$\Tilde{\om}$ on $M$ by  
\be
\Tilde{\om}=\begin{pmatrix}
0&\tau_\mu&0\\
\theta^\nu&\Gamma^\nu_{~\mu}&\eta^{\nu\rho}\tau_\rho\\
0&\eta_{\mu\rho}\theta^\rho&0
\end{pmatrix}.\label{tiom}
\ee
Then, we use a Lie subgroup ${\bf H}$ of ${\bf SO}(2,4)$, generated by the 
$6\times 6$ matrices of the form 
\be
b=\begin{pmatrix}
{\rm e}^{-\phi}&{\rm e}^{-\phi}\xi_\mu&\frac{1}{2}{\rm
  e}^{-\phi}\xi_\mu\xi_\nu \eta^{\mu\nu}\\
&&\\
0&\Lambda^{\nu}_{~\mu}&\Lambda^\nu_{~\rho}\eta^{\rho\sigma}
\xi_\sigma\\
&&\\
0&0&{\rm e}^\phi
\end{pmatrix}, \quad\quad\Lambda^\mu_{~\rho}\Lambda^\nu_{~\sigma}\eta_{\mu\nu}=\eta_{\rho\sigma}, \label{podh}
\ee  
to lift the form $\Tilde{\om}$ to an ${\bf so}(2,4)$-valued 
1-form $\om$ on $M\times H$. Explicitely, if $b$ is a generic element
of ${\bf H}$, we put
$$
\om=b^{-1}\Tilde{\om}b+b^{-1}\der b,
$$
so that 
\beq
&\om=\begin{pmatrix}
-\frac{1}{2}A&\tau'_\mu&0\\
&&\\
\theta'^\nu
&\G'^\nu_{~\mu}&\eta^{\nu\sigma}\tau'_\sigma\label{cc2}\\
&&\\
0&\theta'^\sigma \eta_{\sigma\mu}&
\frac{1}{2}A
\end{pmatrix},\nonumber\\
\eeq
with
\beq
&\theta'^\nu=
{\rm e}^{-\phi}\Lambda^{-1\nu}_{~~~~\rho}\theta^\rho,\\
&\nonumber\\
&A=2\xi_\mu\theta'^\mu+2\der\phi,\nonumber\\
&\nonumber\\
&\G'^\nu_{~\mu}=\Lambda^{-1\nu}_{~~~~\rho}
\Gamma^\rho_{~\sigma}\Lambda^\sigma_{~\mu}+\Lambda^{-1\nu}_{~~~~\rho}
\der\Lambda^\rho_{~\mu}+\theta'^\nu
\xi_\mu-\xi^\nu\eta_{\mu\rho}\theta'^\rho,
\nonumber\\
&\nonumber\\
&\tau'_\mu={\rm
  e}^{\phi}\tau_\nu\Lambda^\nu_{~\mu}-\xi_\alpha
\Lambda^{-1\alpha}_{~~~~\rho}
\Gamma^\rho_{~\nu}\Lambda^\nu_{~\mu}+\frac{1}{2}{\rm
  e}^{-\phi}\eta^{\alpha\beta}\xi_\alpha
\xi_\beta\tau_\nu\Lambda^\nu_{~\mu}-\frac{1}{2}\xi_\mu A+\der \xi_\mu-
\xi_\alpha\Lambda^{-1\alpha}_{~~~~\rho}\der\Lambda^\rho_{~\mu}\nonumber
\eeq 
The so defined form $\om$ is the Cartan normal conformal connection 
associated with the conformal class $[g]$ written in a particular
trivialization of an approppriately defined $H$-bundle over $M$. It
can be viewed as a useful tool for encoding conformal properties of
the metrics on manifolds. Indeed, if given $g$ on $M$ one calculates
the quantities $\theta^\mu,\G^\mu_{~\nu},\tau_\mu$, then the
corresponding quantities for the conformally rescaled metric 
$g'={\rm e}^{-2\phi}g$ are given by (\ref{cc2}) with
$A=0$.\footnote{Note that $A=0$ means that
  $\xi_\mu=-\nabla'_\mu \phi$, where $\nabla'_\mu=
e^\phi \Lambda^\nu_{~\mu}\nabla_\nu$. Admitting $\xi_\mu$ which are not
gradients in the definition of ${\bf H}$ allows for transformations
between different Weyl geometries $[(g,A)]$} \\

\noindent
The curvature of $\om$ is  
$$
R=\der\om+\om\dz\om
$$
and has the rather simple form
\beq
&R=\begin{pmatrix}
0&(D\tau_\mu)'&0\\
&&\\
0&C'^\nu_{~~\mu}
&g^{\nu\mu}(D\tau_\mu)'\\
&&\\
0&0&0
\end{pmatrix}
\eeq
with the 2-forms $C'^\nu_{~~\mu}$ and $(D\tau_\mu)'$ defined by
\beq
&C'^\nu_{~~\mu}=\Lambda^{-1\nu}_{~~~~\rho}C^\rho_{~\sigma}\Lambda^\sigma_{~\mu}\label{trtau}\\
&(D\tau_\mu)'={\rm e}^{\phi}{\rm D}\tau_\nu\Lambda^\nu_{~\mu}-\xi_\alpha
\Lambda^{-1\alpha}_{~~~~\rho}\C^\rho_{~\nu}\Lambda^\nu_{~\mu}.\nonumber
\eeq
Similarly to the properties of $\om$, the curvature $R$ can be
used to extract the transformations of $D\tau_\mu$ and $C^\nu_{~\mu}$
under the conformal rescaling of the metrics. If $g\to g'={\rm
  e}^{-2\phi}g$ these transformations are given by (\ref{trtau}) with
$\xi_\mu=-\nabla'_\mu\phi$. In particular, if we freeze the
Lorentz transformations of the tetrad,
$\Lambda^\mu_{~\nu}=\delta^\mu_{~\nu}$, then we see that the Weyl
2-forms $C^\nu_{~\mu}$ constitute the conformal invariant.\\

\noindent
The curvature $R$ of the Cartan normal connection $\om$ is horizontal
which, in other words, means that it has only 
$\theta^\mu\dz\theta^\nu$ terms in the decomposition onto the basis of
forms
$(\theta^\mu,\der\phi,\Lambda^{-1\mu}_{~~~\nu}\der\Lambda^\nu_{~\rho},\der\xi_\mu)$.
Thus, the Hodge $*$ operator associated with $g$ on $M$ is well
defined acting on $R$ and in consequence the 
Yang-Mills equations for $R$ can be written
\be
D*R=\der *R-*R\dz\om+\om\dz *R=0.
\ee
The following theorem is well known (see e.g. \cite{LK}).
\bt
The metric $g$ on a 4-dimansional manifold $M$
 satisfies the Bach equations $B_{\mu\nu}=0$ if and only if it
 satisfies the Yang-Mills equations $D*R=0$ for the Cartan normal
 conformal connection associated with $g$.
\et
In view of this theorem it is natural to ask about the normal
conformal connection interpretation of $N_{\mu\nu\rho}=0$, which
together with $B_{\mu\nu}=0$ are sufficient for $g$ to be conformal to
Einstein. To answer this question we introduce indices $A,B,C,...$
which run from 0 to 5 and attach them to any $6\times 6$ matrix. 
In this way the elements of matrix $R$ are 2-forms 
\be
R^A_{~B}=\frac{1}{2}R^A_{~B\mu\nu}\theta'^\mu\dz\theta'^\nu.
\ee 
We define the $6\times 6$ matrix of 2-forms $\hat{R}^3$, which is the
appropriately contracted triple product of $R$, with matrix elements 
\be
\hat{R}^3_{AF}=\frac{1}{2}Q_{AE}R^E_{~B\al\beta}R^B_{C\gamma\delta}R^C_{~F\mu\nu}\eta^{\alpha\gamma}\eta^{\beta\delta}\theta'^\mu\dz\theta'^\nu.
\ee 
The symmetric part of this matrix $(\hat{R}^3)^T+\hat{R}^3$ is of the form
\be
(\hat{R}^3)^T+\hat{R}^3=\begin{pmatrix}
0&0&0\\
&&\\
0&0
&P_\mu\\
&&\\
0&P_\mu&P
\end{pmatrix}
\ee 
where $P_\mu$ and $P$ are appropriate 2-forms on
$M\times H$. It follows that under the assumption that
\be
C^2\neq 0
\ee
this matrix has particularly simple form
\be
(\hat{R}^3)^T+\hat{R}^3=\frac{1}{2}{\rm e}^{6\phi}\begin{pmatrix}
0&0&0\\
&&\\
0&0
&{\rm e}^{-\phi}\Lambda^\alpha_{~\sigma}N_{\alpha\beta\gamma}\theta^\beta\dz\theta^\gamma\\
&&\\
0&{\rm
  e}^{-\phi}\Lambda^\alpha_{~\sigma}N_{\alpha\beta\gamma}\theta^\beta\dz\theta^\gamma&V^\alpha N_{\alpha\beta\gamma}\theta^\beta\dz\theta^\gamma
\end{pmatrix}\label{newman}
\ee 
where
\be
V^\alpha=\frac{4}{C^2}(\nabla_\rho
C^\rho_{~\lambda\tau\sigma})C^{\alpha\lambda\tau\sigma}-{\rm
  e}^{-\phi}\xi_\rho\Lambda^{-1\rho}_{~~~~\lambda}\eta^{\lambda\alpha}
\ee
and $N_{\alpha\beta\gamma}$ is given by (\ref{ted}).
The proof of this fact consists of a straightforward but lengthy
calculation which uses the identities (\ref{wid}) and (\ref{for}).
This enables us to formulate the following theorem. 
\bt
Assume that the metric $g$ satisfies $$C^2\neq 0$$ on $M$. Let
$\omega$ be its Cartan normal conformal connection with curvature
$R$ and the matrix $\hat{R}^3$ as above. Then the
metric is locally conformally equivalent to the Einstein metric if and
only if the 
\be
{\rm (i)}~~~~D*R=0~~~~~~~~~{\rm and}~~~~~~~~~{\rm (ii)}~~~~(\hat{R}^3)^T+\hat{R}^3=0.
\ee
\et
\phantom\\
\noindent
The above condition (ii) can now be compared to the Baston-Mason conformal
connection interpretation of the condition
$E_{\mu\nu\rho}=0$. According to them, this condition is \cite{Mas}
\be
{\rm (ii')}~~~~~~~~~~~~~~[R^+_{~\mu\nu}, R^-_{\rho\sigma}]=0,
\ee
where $R^+=\frac{1}{2}R^+_{~\mu\nu}\theta^\mu\dz\theta^\nu$ and
$R^-=\frac{1}{2}R^-_{~\mu\nu}\theta^\mu\dz\theta^\nu$ denote, 
respectively, the self-dual and anti-self-dual
parts of the curvature $R$, i.e. $*R^\pm=\pm i R^\pm$, and $R=R^+\oplus
R^-$.\\

\noindent
{\bf Acknowledgments}\\
We acknowledge support from NSF Grant No PHY-0088951 and the Polish
KBN Grant No 2 P03B 12724. 

\end{document}